\documentclass{aastex}          %% The manuscript based on AASTeX v5.x
\usepackage{spr-astr-addons}    %% mimicing ASTR journal style
\usepackage{nccmath}
\usepackage{xcolor}

\begin{document}
%% Article title
%
\title{Evidence for possible systematic underestimation of uncertainties in extragalactic distances and its cosmological implications}

%% Running heads
\shorttitle{Underestimation of uncertainties in extragalactic distances}
\shortauthors{Ritesh Singh}

%% Author and Affilations
\author{Ritesh Singh\altaffilmark{}} 
%\and 
%\author{\altaffilmark{}}
\affil{Samsung Research Institute, India}
\email{ritesh.s7@samsung.com} %% non-output

%% Alternate Affilations
%\altaffiltext{1}{Affilation}
%\altaffiltext{2}{}
%\altaffiltext{3}{}

%% Abstract
\begin{abstract}
We analyze 91,742 reported extragalactic distance moduli and their one sigma uncertainties for 14,560 galaxies with multiple reported distances in the NED Redshift-Independent Distances database. For every ordered pair of distance moduli measurements 1 and 2 for each galaxy, we define $\Delta(\sigma)_{1,2}$, a measure for how different measurement 2 is in relation to measurement 1, as a multiple of the reported one sigma uncertainty in measurement 1. For a given set of distance moduli measurements, we take a mean of all such $\Delta(\sigma)_{1,2}$ to determine the average separation between distance moduli of any galaxy in the set as a multiple of the reported uncertainty. For normally distributed measurements, the expected value of mean $\Delta(\sigma)$ is 0.79. Our results are as follows. The mean $\Delta(\sigma)$ of 1,239,062 ordered pairs of extragalactic distance moduli for 14,560 galaxies is 2.07 corresponding to a p-value of 3.85\%. This indicates a possible systematic underestimation of uncertainties in extragalactic distances. We also find that the mean reported one sigma uncertainty decreased and the mean $\Delta(\sigma)$ increased from 2004 to 2018. This points to increased underestimation of uncertainties with time. For the latest period from 2014 to 2018 with 14,580 reported extragalactic distance moduli for 5,406 galaxies, the mean $\Delta(\sigma)$ of 40,462 ordered pairs is 3.00 corresponding to a p-value of 0.27\%. For 14,888 extragalactic distance moduli of 2,518 galaxies measured using Type Ia Supernovae, the mean $\Delta(\sigma)$ for 124,016 ordered pairs is 2.85 corresponding to a p-value of 0.44\%. These results may have some implications for our confidence in cosmological parameters and models. We conclude that more liberal estimation of uncertainties in future reported extragalactic distances should be considered. The results also give a possible way out of the Hubble-Lemaitre tension by advocating for increasing the error bars on Hubble-Lemaitre constant measured via distance ladders of standard candles and rulers.
\end{abstract}

%% Keywords

%%  Please use labels (\label, \ref) for section, figure, table, 
%%  equation  reference. Use \cite for bibliography references.
%
\section{Highlights} %\label{s:?}
\begin{enumerate}

\item Accurate measurement of extragalactic distances and estimation of associated uncertainties directly determines our confidence in cosmological parameters and models.

\item We find that any two distance moduli measurements of the same galaxy differ from each other by 2.07 times the reported one sigma uncertainty on average.

\item This average difference between distance moduli measurements of the same galaxy as a multiple of reported uncertainty is growing with time of publication, rising to 3.00 times the reported one sigma uncertainty for all distances reported from 2014 to 2018.

\item This average difference between distance moduli measurements of the same galaxy as a multiple of reported one sigma uncertainty is highest for the standard candles (3.01) including Cepheids (4.26), Type Ia Supernovae (2.85), and Tip of the Red Giant Branch (2.82).

\item This data points to a possible systematic underestimation of uncertainties in extragalactic distances.

\item The results also give a possible way out of the Hubble-Lemaitre tension by advocating for increasing the error bars on Hubble-Lemaitre constant measured via distance ladders of standard candles and rulers.

\end{enumerate}

\section{Introduction}
Extragalactic distances are critical pillars of cosmological models. Accurate measurement of extragalactic distances and estimation of associated uncertainties directly determines our confidence in cosmological parameters and models \citep{czerny_astronomical_2018}. It also has direct implications for our understanding of the local structure and the large scale structure of the Universe, and even transient gravitational wave detections \citep{chaparro-molano_predicting_2019}.

The only independently determined distances in astronomy are the trigonometric parallaxes which can only be accurately estimated for small distances up to about 1000 parsecs. All other methods of distance determination are calibrated with data from known distances from other methods, starting from parallax distances \citep{krisciunas_first_2012}. As an example, parallax distances up to about one Kpc were used to calibrate the Cepheid distance scale useful up to a few Mpc, which was in turn used to calibrate the Type Ia Supernovae scale useful up to 100 Mpc and beyond. Such calibration of increasingly farther distance indicators with hitherto established distance measurement methods extends the reach of our scale immensely. This sequence of distance indicators with increasing reach is known as the cosmic distance ladder or the extragalactic distance scale \citep{rowan-robinson_cosmological_1985, rowan-robinson_extragalactic_1988, rowan-robinson_climbing_2008}.

This interdependence of distance estimation methods leads to propagation of uncertainties. Therefore, if the uncertainties in a portion of the distance measurements are underestimated, as we show that they could be, the underestimation itself is prone to a system wide propagation.

While the measurement of extragalactic distances is itself an inexact science, the estimation of uncertainties in the distances is even more approximate and prone to errors \citep{rowan-robinson_type_2002}. This is in part due to the interdependence of the distance measurement methods on other data and models, which themselves are empirical approximations.

In spite of this, the reported confidence in measured extragalactic distances has gradually increased with time due to improvements in theory, quality and quantity of data, and statistical methods \citep{chaparro-molano_predicting_2019}.

Extragalactic distance databases compile distance measurements of the same galaxy using various different methods and/or by different teams, for numerous galaxies. They offer a unique opportunity for comparative analysis, which can grant an insight into the relative accuracy and precision of the different methods, as well as a way to measure the adequacy of the reported uncertainties in the measured extragalactic distances.

One such extragalactic distance database is the NED Redshift-Independent Distances database \citep{steer_redshift-independent_2017}. For 14,560 galaxies, it collates multiple reported distances making a total of 91,742 distances. If the measured distances of the same galaxy by different methods and/or teams lie well beyond the reported one sigma uncertainties, we can conclude that the reported uncertainties are inadequate. We attempt to carry out such an analysis in this paper using the NED database.

The fact that the reported uncertainty in extragalactic distances is much less than the scatter across multiple distance measurements for the same galaxy is also noted by \cite{chaparro-molano_predicting_2019}. Much earlier, \cite{jacoby_critical_1992} termed the underestimation of uncertainties in extragalactic distances to be a ‘major embarrassment’, while \cite{rowan-robinson_cosmological_1985,rowan-robinson_extragalactic_1988} called it ‘incredible’.

Apart from widespread implications for various cosmological theories and models, the increasing confidence in extragalactic distances has precipitated a dichotomy in the values of the Hubble-Lemaitre constant \citep{jackson_hubble_2015}. The larger than expected discrepancy between values of the Hubble-Lemaitre constant, among those measured by various seemingly independent methods, has been called as a ‘crisis’ or ‘tension’ in Cosmology \citep{di_valentino_planck_2020,crane_cosmological_2019, chen_hubble_2018,gonzalez_crisis_2019,chen_sharp_2019,riess_large_2019}.

In particular, the sharp difference (relative to the reported uncertainties) between the values of the Hubble-Lemaitre constant estimated from observations of the Cosmic Microwave Background (CMB), and those estimated by Type Ia Supernovae observations and observations based on Tip of the Red Giant Branch (TRGB), is cause for much intrigue recently in the academia, literature, and science news \citep{rameez_is_2021,sloman_hubble_2019,poulin_hubble_2019,ivanov_h_2020,di_valentino_realm_2021,sivaram_hubble_2021,krishnan_running_2021}. This difference in the estimated values of the Hubble-Lemaitre constant is neither a recent phenomenon, nor is it strange by itself \citep{rowan-robinson_climbing_2008}. However, with the increasing confidence in the extragalactic distances, the different estimates have gone well beyond the reported uncertainties \citep{jackson_hubble_2015}. 

If it is established that the uncertainties in extragalactic distances are indeed underestimated as it seems, the error bars on the Hubble-Lemaitre constant determined by the measurement of extragalactic distances and recession velocities needs to be increased. If this is done, the tension withers away.

\section{Data}

We analyze 91,742 reported extragalactic distance moduli and their one sigma uncertainties for 14,560 galaxies with multiple reported distances in the NED Redshift-Independent Distances database. \cite{steer_redshift-independent_2017} have meticulously compiled the reported distances from those published in the SAO/NASA Astrophysics Data System \citep{kurtz_nasa_2000}. The nearest analyzed galaxy is 2FGL J1507.0-6223 at a distance of 1.7 Kpc and the furthest is SDSS-II SN 13151 at a distance of 94.5 Gpc. We exclude the Large and the Small Clouds of Magellan since their size is large in proportion to their distance. We also exclude the reported distances with no reported uncertainty: 13,350 values out of the total 328,317 values enlisted in the NED database.

The NED database includes the following distance measurement methods \citep{steer_redshift-independent_2017}:

\textbf{Standard Candles:} AGN time lag, Asymptotic Giant Branch Stars, BL Lac Object Luminosity, Black Hole, Brightest Cluster Galaxy, Brightest Stars, Carbon Stars, Cepheids, Color-Magnitude Diagrams, Delta Scuti, Flux-Weighted Gravity-Luminosity Relation, Gamma-Ray Burst, Globular Cluster Luminosity Function, Globular Cluster Surface Brightness Fluctuations, HII Luminosity Function, Horizontal Branch, M Stars luminosity, Miras, Novae, Planetary Nebula Luminosity Function, RR Lyrae Stars, Red Clump, Red Supergiant Variables, SNIa SDSS, SX Phoenicis Stars, Short Gamma-Ray Bursts, Statistical, Sunyaev-Zeldovich Effect, Surface Brightness Fluctuations, Tip of the Red Giant Branch, Type II Cepheids, Type II Supernovae Radio, Type Ia Supernovae, Wolf-Rayet.

\textbf{Standard Rulers:} Dwarf Galaxy Diameter, Eclipsing Binary, Globular Cluster Radii, Grav. Stability Gas. Disk, Gravitational Lenses, HII Region Diameters, Jet Proper Motion, Masers, Orbital Mechanics, Proper Motion, Ring Diameter, Type II Supernovae Optical.

\textbf{Secondary Methods:} D-Sigma, Diameter, Faber-Jackson, Fundamental Plane, GeV TeV ratio, Globular Cluster Fundamental Plane, Gravitational Wave, H I + optical distribution, Infra-Red Astronomical Satellite, LSB galaxies, Magnitude, Mass Model, Sosies, Tertiary, Tully Estimate, Tully-Fisher.

About 20\% of the reported distance moduli are on a non-standard extragalactic distance scale, meaning that either they assume the value of the Hubble-Lemaitre constant to be different from 70 km/s/Mpc, or they assume the Large Magellanic Cloud zero point distance modulus to be different from 18.50 \citep{2019AAS...23340901S}. To be compared on the same scale, such distance moduli are scaled to the standard distance scale with $H_0=$ 70 km/s/Mpc [used by the Supernova Cosmology Project, the Supernova Legacy Survey, and others (Steer et al. 2017)] and the Large Magellanic Cloud zero point distance modulus equal to 18.50. The reported percentage uncertainty is retained. Where luminosity distance moduli are reported, they are converted into proper distance moduli.

\section{Analysis}

We analyze extragalactic distance moduli measurements and their reported one sigma uncertainties by grouping them into sets. Some such sets of distance moduli are based on distance measurement methods, while others are based on period of publication. These sets contain distance moduli measurements of multiple galaxies, with multiple distance moduli for each galaxy.

For all extragalactic distance moduli measurements in a given set, we calculate mean $\sigma$ for that set by taking a mean of all the reported one sigma uncertainties in percentage.

Furthermore, for every ordered pair of distance moduli measurements 1 and 2 for each galaxy, we calculate $\Delta(\sigma)_{1,2}$, a measure for how different measurement 2 ($m_2$) is in relation to measurement 1 ($m_1$), as a multiple of the reported one sigma uncertainty in measurement 1 ($\sigma_1$). In other words, we define $\Delta(\sigma)_{1,2}$ as follows:

\begin{ceqn}
\begin{equation}
{\Delta(\sigma)_{1,2}=\dfrac{\left|m_1 - m_2\right|}{\sigma_1}}
\end{equation}
\end{ceqn}

For a given set of distance moduli measurements, we take a mean of all such $\Delta(\sigma)_{1,2}$ to determine the average separation between distance moduli of any galaxy in the set as a multiple of the reported uncertainty. In other words, the mean $\Delta(\sigma)$ of a set of distance moduli measurements is the expected separation of any two distance moduli in the set for any particular galaxy, as a multiple of the uncertainty reported in any one of them.

Such a grouping of distance moduli measurements in a set, and calculating mean $\sigma$ and mean $\Delta(\sigma)$ for the set, enables us to identify the distance measurement methods which show greater underestimation of uncertainties. It also helps us to investigate trends in estimation of uncertainties over the decades. Both these questions are investigated in detail in this paper.

It is important to note that no inter-set pairing is considered while calculating mean $\Delta(\sigma)$ for a set. This is done to avoid inaccuracies and biases in other sets from affecting the mean $\Delta(\sigma)$ of the set.

If measurement 1 is assumed to be normally distributed with mean $m_1$ and standard deviation $\sigma_1$, the expected value of $\Delta(\sigma)_{1,2}$ is just the mean absolute deviation of a normal distribution divided by its standard deviation, which is $(\sigma\sqrt{2/\pi})/\sigma=0.79$. A value of mean $\Delta(\sigma)$ significantly larger than 0.79 points to a possible underestimation of uncertainties.

Another value may also hold some statistical relevance for the purpose of comparison. The mean absolute separation of any two independent measurements sampled from a normal distribution is just the mean of a folded normal distribution corresponding to a normal distribution with mean 0 and variance $2\sigma^2$, which is $2\sigma/\sqrt{\pi}=1.13\sigma$. In other words, if $m_1$ and $m_2$ were independent measurements of a global normal distribution with standard deviation $\sigma_g$, then the expected value of $\left|m_1 - m_2\right|$ would be $1.13\sigma_g$. 

However, $m_1$ is not reported to be just another independent measurement of a global normal distribution. On the contrary, when $m_1$ is reported, what is actually reported is that the distance moduli measurements to that galaxy are expected to be normally distributed with mean $m_1$ and standard deviation $\sigma_1$. Therefore, we should expect the mean deviation of other measurements from $m_1$ to be $0.79\sigma_1$. In other words, in order to be consistent with other reported measurements, $\sigma_1$ needs to be larger if $m_1$ is the reported mean, than when $m_1$ is just one of the reported independent measurements of a global normal distribution. 

Therefore, in our opinion, comparing mean $\Delta(\sigma)$ with 0.79 gives a better measure of underestimation of reported uncertainty in distance moduli measurements. However, if the reader prefers to compare it with 1.13, they may choose to do so. In any case, we have only highlighted the distance measurement methods where the mean $\Delta(\sigma)$ is greater than 2. Besides, this choice of comparison metric has no effect on our reported values in this paper.

It may be noted that calculating $\Delta(\sigma)_{1,2}$ using square of the difference between measurements $(m_1-m_2)^2$, rather than the absolute difference $\left|m_1 - m_2\right|$, would have enabled a direct comparison with the standard deviation $\sigma$ instead of the mean absolute deviation $0.79\sigma$. However, standard deviation, owing to squaring of the difference from the mean, weighs larger deviations more heavily as compared to smaller ones. This is unnecessary for our purposes since we do not want outliers to skew the mean $\Delta(\sigma)$, which would magnify the problem more than what may be considered correct.

The mean $\Delta(\sigma)$ for each distance measurement method is also used to calculate the p-value. The p-value gives the probability of measuring a value differing from the mean by more than mean $\Delta(\sigma)$ times the standard deviation, for a normal distribution. The essence of the p-value is the likelihood of finding a measurement far away from mean. It means that, if the reported one sigma uncertainties  were accurately estimated, the likelihood of finding a measurement more than mean $\Delta(\sigma)$ away from the mean would be the p-value. 

\section{Results and Discussion}

We create all possible ordered pairs of distance moduli measurements for each galaxy. Doing this for 14,560 galaxies with multiple reported distance measurements in the NED database makes a total of 1,239,062 ordered pairs.

The mean reported one sigma uncertainty of all 91,742 reported extragalactic distance moduli for 14,560 galaxies is 1.09\%. The mean $\Delta(\sigma)$ of all 1,239,062 ordered pairs of extragalactic distance moduli is 2.07 corresponding to a p-value of 3.85\%. 

In other words, for a given pair of reported distance moduli measurements for any particular galaxy, one measurement is expected to differ from the other by $2.07\sigma$, or 2.07 times the reported one sigma uncertainty in it. The p-value implies that the probability of this is only 3.85\% for normally distributed measurements, meaning that only a small minority of measurements should differ by more than 2.07 times the reported one sigma uncertainty; however, in our case the measurements differ by $2.07\sigma$ on average.

Thus, the mean $\Delta(\sigma)$ being significantly high (and the p-value being significantly low), indicates a possible systematic underestimation of uncertainties in extragalactic distances.

\begin{figure*}
 \centering\includegraphics[width=0.75\textwidth]{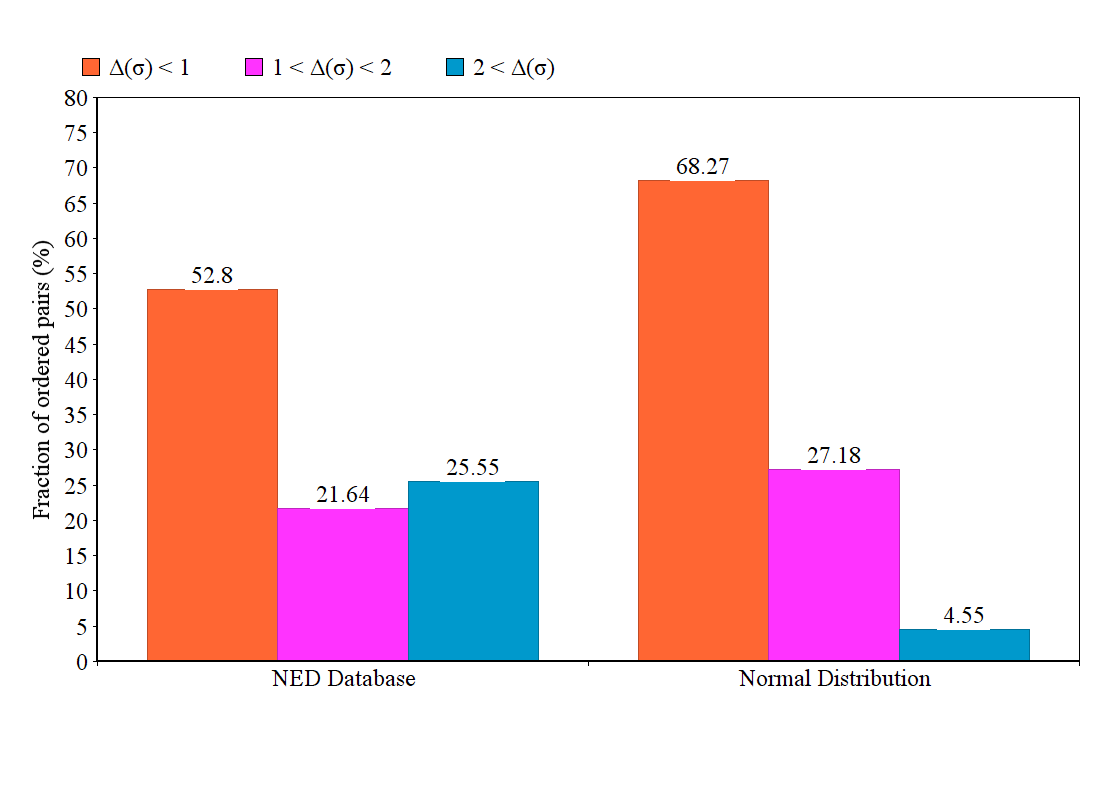}
 \caption{Percentage of 1,239,062 ordered pairs of extragalactic distance moduli over ranges of $\Delta(\sigma)$ compared with a normal distribution. A large fraction of the ordered pairs differ by a significantly large multiple of their reported uncertainties [$\Delta(\sigma)>2$], pointing to a possible underestimation of the uncertainties} %% no full stop at the end of caption
 \label{fig:1}
 \end{figure*}

Fig. \ref{fig:1} shows the percentage of 1,239,062 ordered pairs of extragalactic distance moduli over ranges of $\Delta(\sigma)$. It also shows the expected percentage for normally distributed measurements. It is a standard result in statistics that for a normal distribution, 68.27\% of the values lie within one standard deviation [that is $\Delta(\sigma)<1$], 27.18\% of the values lie between one and two standard deviations [that is $1<\Delta(\sigma)<2$], and 4.55\% values lie beyond two standard deviations [that is $2<\Delta(\sigma)$].
 
We see that 25.55\% of the ordered pairs differ by a significantly large multiple of their reported uncertainties [$\Delta(\sigma)>2$]. The fraction is much greater than the expected percentage for normally distributed measurements (4.55\%), pointing towards a possible underestimation of the uncertainties.

\subsection{Mean $\sigma$, mean $\Delta(\sigma)$, and p-value for different distance measurement methods}

The NED database includes three categories of distance measurement methods: Standard Candles, Standard Rulers, and Secondary Methods.

The mean reported one sigma uncertainty, the mean $\Delta(\sigma)$, and the corresponding p-value for each of the three categories is given in Table \ref{tbl:1}. The mean $\Delta(\sigma)$ is significantly large (and the p-value significantly small) for both Standard Candles and Standard Rulers, pointing to a possible underestimation of uncertainties in both the categories.

%% Table (two-column)
%
 \begin{table*}%%[tb]
 \small
 \caption{Mean reported one sigma uncertainty, mean $\Delta(\sigma)$, and corresponding p-value for three categories of distance measurement methods} %% no full stop at the end of caption
 \label{tbl:1}
 \begin{tabular}{c|c|c|c|c|c|c}
 \tableline  %% rule at top
Category & Mean $\sigma$ (\%) & No. of observations & Mean $\Delta(\sigma)$ & No. of pairs & No. of galaxies & p-value (\%) \\ \hline
Standard Candles & 0.57 & 25184 & \color{red} 3.01 & 424952 & 4758 & \color{red} 0.26 \\
Standard Rulers & 0.99 & 983 & \color{red} 2.31 & 5708 & 235 & \color{red} 2.10 \\
Secondary Methods & 1.29 & 64850 & 0.99 & 613766 & 10174 & 32.22 \\
% \tablenotemark{a} 
% <entries>
 \tableline %% rule at bottom
 \end{tabular}
%
%% Any table notes must follow the \end{tabular} command. 
% \tablenotetext{a}{}
 \tablecomments{The mean $\Delta(\sigma)$ is too large and the p-value too small for both the Standard Candles and the Standard Rulers (colored red), pointing to a possible underestimation of uncertainties in those categories.}
 \end{table*}

\begin{figure*}
 \centering\includegraphics[width=0.75\textwidth]{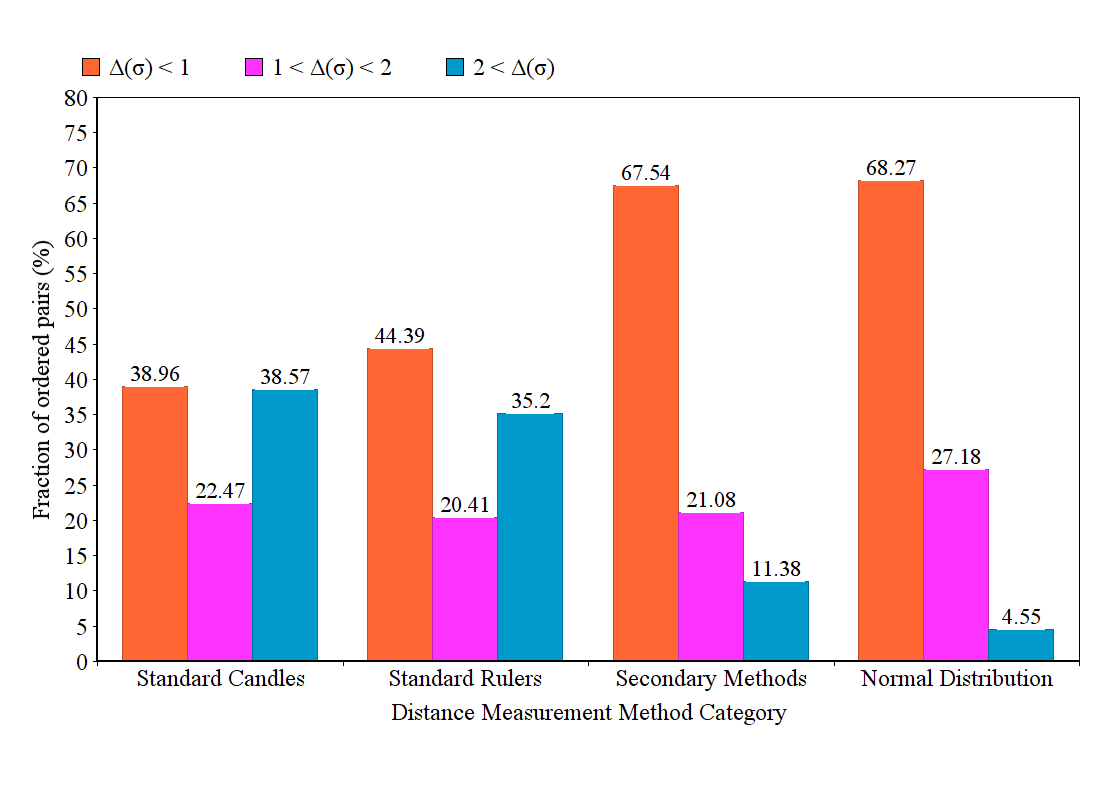}
 \caption{Percentage of ordered pairs  of extragalactic distance moduli over ranges of $\Delta(\sigma)$ for three categories of distance measurement methods, compared with a normal distribution. A large fraction of the ordered pairs differ by a significantly large multiple of their reported uncertainties [$\Delta(\sigma)>2$] for both the Standard Candles and the Standard Rulers, pointing to a possible underestimation of uncertainties in those categories} %% no full stop at the end of caption
 \label{fig:2}
 \end{figure*}

Fig. \ref{fig:2} shows the percentage of ordered pairs of extragalaxtic distance moduli over ranges of $\Delta(\sigma)$ for each of the three categories, compared with a normal distribution.

A large fraction of the ordered pairs differ by a significantly large multiple of their reported uncertainties [$\Delta(\sigma)>2$] for both the Standard Candles (38.57\%) and the Standard Rulers (35.20\%). The fractions are much greater than the expected percentage for normally distributed measurements (4.55\%), pointing to a possible underestimation of uncertainties in those categories.

Since both the standard candles and the standard rulers are used to construct the cosmic distance ladder, underestimation of their uncertainties has some implications for our confidence in cosmological parameters and models.

While the Standard Candles and the Standard Rulers are deemed to be more accurate than the Secondary Methods (as is also clear from their mean reported one sigma uncertainty of 0.57\% and 0.99\% respectively, as compared to 1.29\% for the Secondary Methods), these results show that at least some of that greater accuracy could be only an artifact of underestimation of uncertainties in those distance measurement methods. 

 The NED database lists 44 types of distance measurement methods with multiple reported measurements for at least one galaxy \citep{steer_redshift-independent_2017}. The mean reported one sigma uncertainty as well as the mean $\Delta(\sigma)$ for each of the methods is given in Table \ref{tbl:3} in the appendix. Table \ref{tbl:4} in the appendix enlists the percentage of ordered pairs of extragalactic distance moduli over ranges of $\Delta(\sigma)$ for each of the 44 distance measurement methods individually.

\begin{figure*}
 \centering\includegraphics[width=0.75\textwidth]{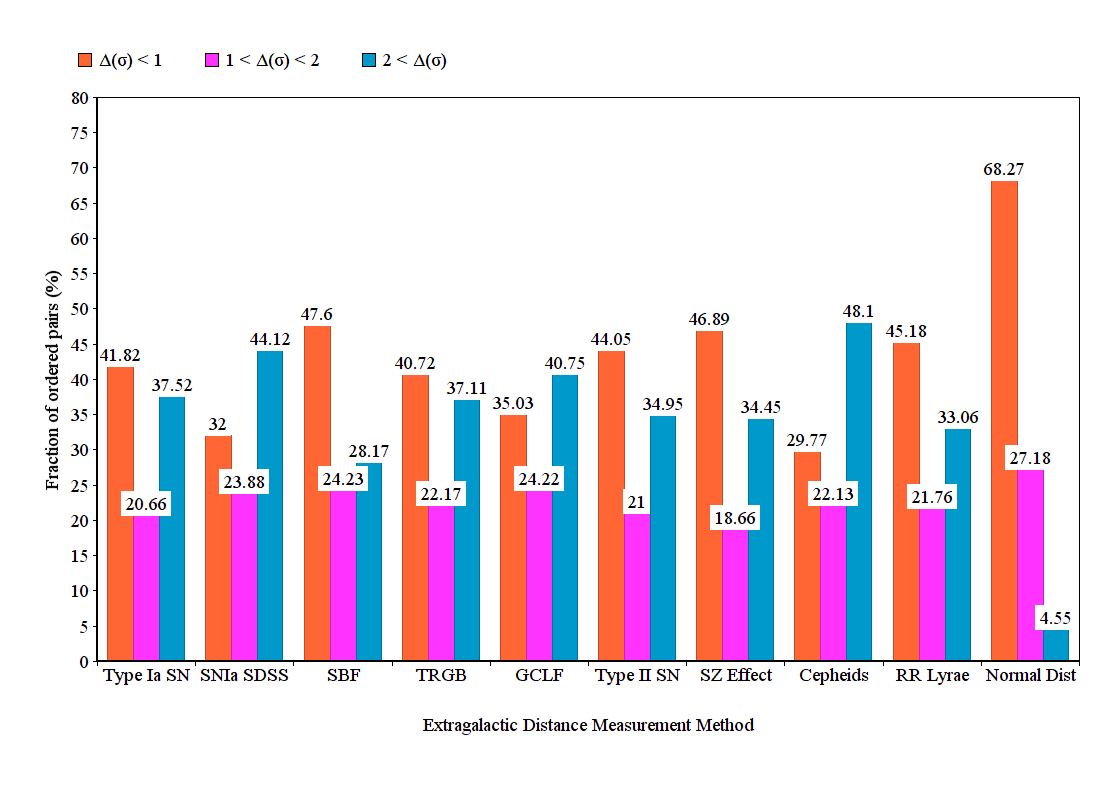}
 \caption{Percentage of ordered pairs of extragalactic distance moduli over ranges of $\Delta(\sigma)$ for 9 distance measurement methods that have $\Delta(\sigma)>2$ for more than 25\% of the ordered pairs (which points to a possible underestimation of uncertainties in those methods). The expected percentage for a normal distribution is also shown} %% no full stop at the end of caption
 \label{fig:3}
 \end{figure*}

Fig. \ref{fig:3} shows the percentage of ordered pairs of extragalactic distance moduli over ranges of $\Delta(\sigma)$ for 9 of the 44 distance measurement methods: those methods that give distance moduli measurements for more than 40 galaxies and have $\Delta(\sigma)>2$ for more than 25\% of the ordered pairs (which points to a possible underestimation of uncertainties in those methods). 

The distance measurement methods plotted in Fig. \ref{fig:3} are Type Ia Supernovae, SNIa SDSS, Surface Brightness Fluctuations, Tip of the Red Giant Branch, Globular Cluster Luminosity Function, Type II Supernovae Optical, Sunyaev-Zeldovich Effect, Cephieds, and RR Lyrae stars. The method names are taken from the NED database \citep{steer_redshift-independent_2017}. The expected percentage for a normal distribution is also shown.

\begin{figure*}
 \centering\includegraphics[width=0.75\textwidth]{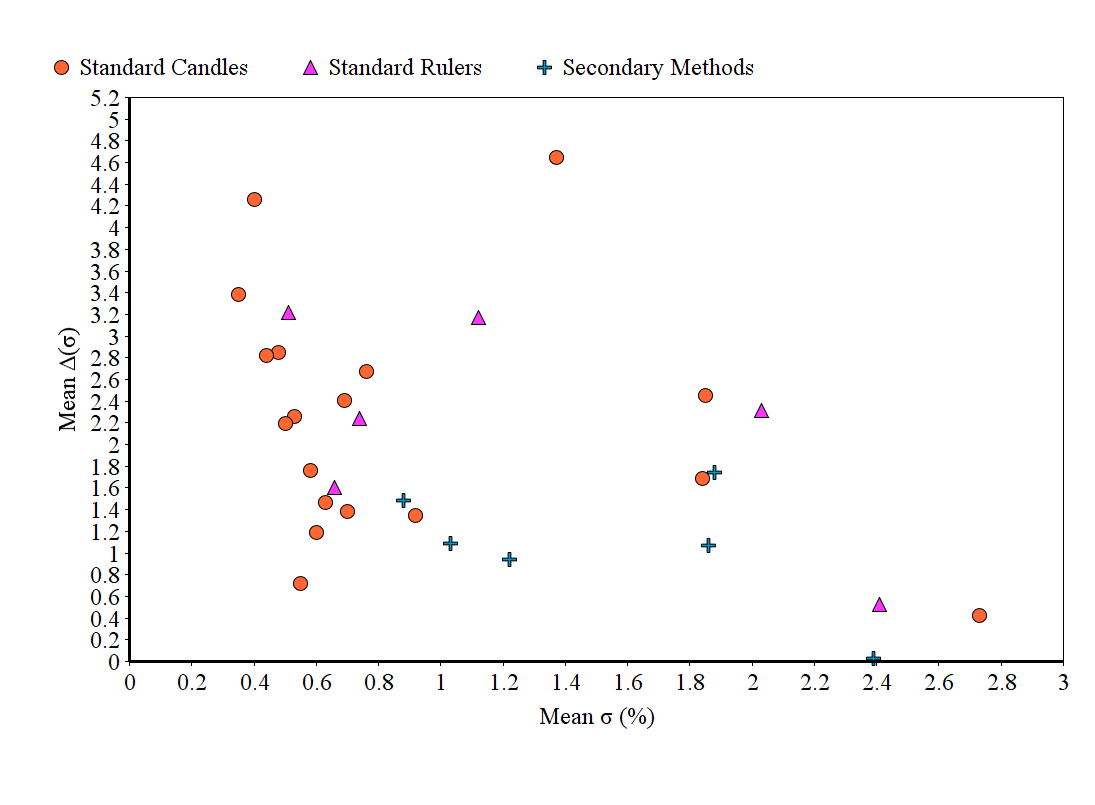}
 \caption{Mean $\Delta(\sigma)$ versus mean reported one sigma uncertainty for 30 distance measurement methods with at least 16 ordered pairs of reported distance moduli. Different categories of methods are represented differently. As expected, a trend of decreasing mean $\Delta(\sigma)$ with increasing mean reported one sigma uncertainty for each category of distance measurement method, as well as across all methods, is seen} %% no full stop at the end of caption
 \label{fig:4}
 \end{figure*}

 Fig. \ref{fig:4} shows the variation of mean $\Delta(\sigma)$ with mean reported one sigma uncertainty for 30 distance measurement methods with at least 16 ordered pairs of reported distance moduli. It shows a trend of decreasing mean $\Delta(\sigma)$ with increasing mean reported one sigma uncertainty for each category of distance measurement method, as well as across all methods, as expected. 
 
 The distance measurement methods plotted in Fig. \ref{fig:4} are Tully-Fisher, Type Ia Supernovae, Cepheids, Tip of the Red Giant Branch, Color-Magnitude Diagrams, Surface Brightness Fluctuations, D-Sigma, Type II Supernovae Optical, Globular Cluster Luminosity Function, Faber-Jackson, RR Lyrae Stars, Fundamental Plane, Gamma-Ray Burst, SNIa SDSS, Sunyaev-Zeldovich Effect, Infra-Red Astronomical Satellite, Planetary Nebula Luminosity Function, Carbon Stars, Brightest Stars, Horizontal Branch, Red Clump, Gravitational Lenses, Type II Cepheids, Ring Diameter, Eclipsing Binary, Statistical, Masers, HII Region Diameters, Miras, GeV TeV ratio \citep{steer_redshift-independent_2017}. Different categories of distance measurement methods are represented differently in the graph.
 
 \subsubsection{Mean $\sigma$, mean $\Delta(\sigma)$, and p-value for distances measured using Type Ia Supernovae}
 
 For 14,888 extragalactic distance moduli of 2,518 galaxies measured using Type Ia Supernovae, the mean reported one sigma uncertainty is 0.48\%, and the mean $\Delta(\sigma)$ for all 124,016 ordered pairs is 2.85 which is significantly high, corresponding to a p-value of 0.44\% which is significantly low, pointing towards possible underestimation of uncertainties.

For distances measured using Type Ia Supernovae, 38.16\% of the ordered pairs differ by a significantly large multiple of their reported uncertainties [$\Delta(\sigma)>2$] against an expected 4.55\% for normal distributions, indicating a possible underestimation of the uncertainties. 

Since Type Ia Supernovae distances form a critical basis of cosmological models and expansion history, underestimation of uncertainties in them may have some implications for our confidence in cosmological parameters and models.

\subsection{Mean $\sigma$, mean $\Delta(\sigma)$, and p-value for different periods of publication (trend with time)}

\begin{figure*}
 \centering\includegraphics[width=0.75\textwidth]{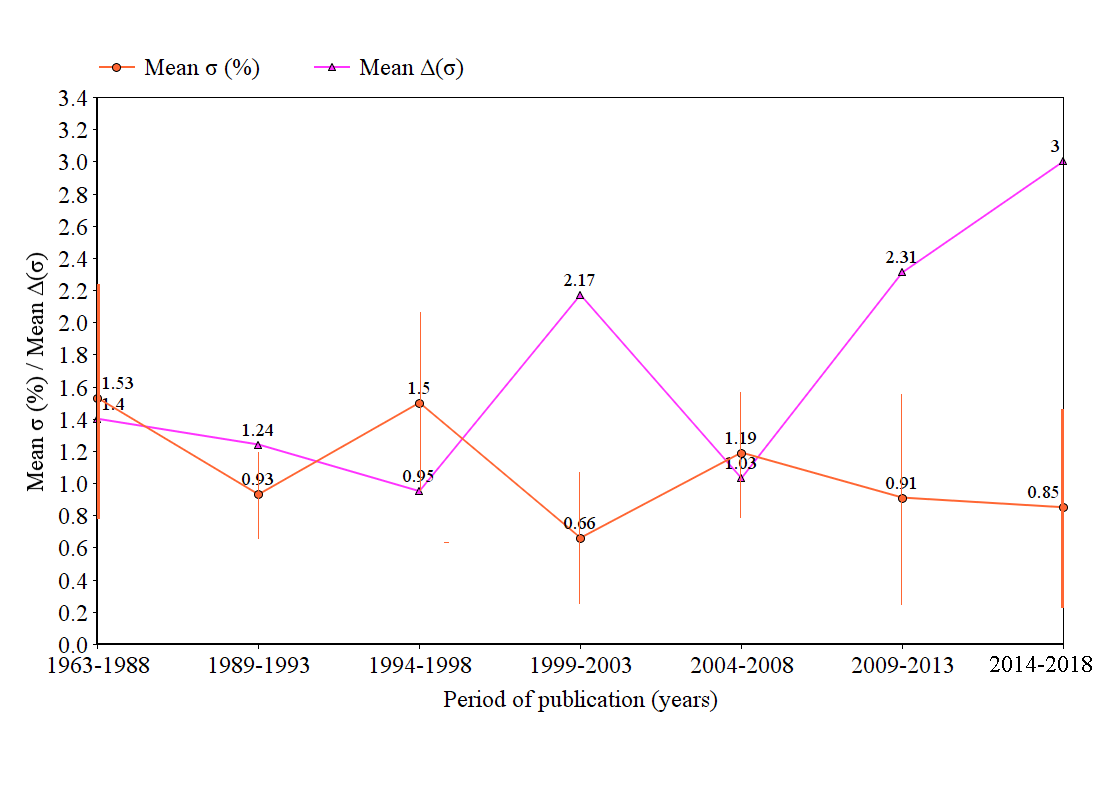}
 \caption{Mean reported one sigma uncertainty with error bars, and mean $\Delta(\sigma)$, for different periods of publication from 1963 to 2018. The mean reported one sigma uncertainty decreased from 2004 to 2018, and the mean $\Delta(\sigma)$ increased in that period. This points to increased underestimation of uncertainties with time} %% no full stop at the end of caption
 \label{fig:5}
 \end{figure*}

The extragalactic distances in the NED database were reported from 1963 to 2018 \citep{2019AAS...23340901S}. In Table \ref{tbl:2} and Fig. \ref{fig:5}, we look at how the mean reported one sigma uncertainty and the mean $\Delta(\sigma)$ varies with the time of publication of the distance measurement. For this, we divide the distance measurements into seven sets based on their year of publication. The periods are 1963 to 1988, 1989 to 1993, 1994 to 1998, 1999 to 2003, 2004 to 2008, 2009 to 2013, and 2014 to 2018.

Fig. \ref{fig:5} shows the mean reported one sigma uncertainty and the mean $\Delta(\sigma)$ for each of the seven periods. It can be seen that, the mean reported one sigma uncertainty and the mean $\Delta(\sigma)$ are inversely correlated with the Pearson Correlation Coefficient $r=-0.68$. Observe also that, as the mean reported one sigma uncertainty decreased from 2004 to 2018, the mean $\Delta(\sigma)$ increased in that period. This points to increased underestimation of uncertainties with time. An important contributing factor in this could be the pressure of improving upon existing uncertainties, as well as peer pressure from other reported measurements.

 Further analysis data for each of the seven periods is given in Table \ref{tbl:2}. For the latest period from 2014 to 2018, the mean reported one sigma uncertainty of 14,580 extragalactic distance moduli for 5,406 galaxies is 0.85\%. The mean $\Delta(\sigma)$ of 40,462 ordered pairs of extragalactic distance moduli reported from 2014 to 2018 is 3.00 which is significantly high, corresponding to a p-value of 0.27\% which is significantly low, pointing to a possible underestimation of uncertainties in that period.
 
 %% Table (two-column)
%
 \begin{table*}%%[tb]
 \small
 \caption{Mean reported one sigma uncertainty, mean $\Delta(\sigma)$, and corresponding p-value for different periods of publication from 1963 to 2018} %% no full stop at the end of caption
 \label{tbl:2}
 \begin{tabular}{c|c|c|c|c|c|c}
 \tableline  %% rule at top
Period & Mean $\sigma$ (\%) & No. of observations & Mean $\Delta(\sigma)$ & No. of pairs & No. of galaxies & p-value (\%) \\ \hline
1963-1998 & 1.53 & 5747 & 1.40 & 28580 & 1213 & 16.15 \\
1989-1993 & 0.93 & 3926 & 1.24 & 6266 & 1807 & 21.50 \\
1994-1998 & 1.46 & 13967 & 0.95 & 66606 & 3142 & 34.21 \\
1999-2003 & 0.66 & 2747 & \color{red} 2.17 & 29714 & 448 & \color{red} 3.00 \\
2004-2008 & 1.19 & 17052 & 1.03 & 56362 & 4838 & 30.30 \\ 
2009-2013 & 0.91 & 27015 & \color{red} 2.31 & 111372 & 7509 & \color{red} 2.09 \\
2014-2018 & 0.85 & 14580 & \color{red} 3.00 & 40462 & 5406 & \color{red} 0.27 \\
% \tablenotemark{a} 
% <entries>
 \tableline %% rule at bottom
 \end{tabular}
%
%% Any table notes must follow the \end{tabular} command. 
% \tablenotetext{a}{}
 \tablecomments{The mean $\Delta(\sigma)$ is too large and the p-value too small for the latest two periods from 2009 to 2013 and 2014 to 2018 (colored red), pointing to a possible underestimation of uncertainties in those periods.}
 \end{table*}

\begin{figure*}
\centering\includegraphics[width=0.75\textwidth]{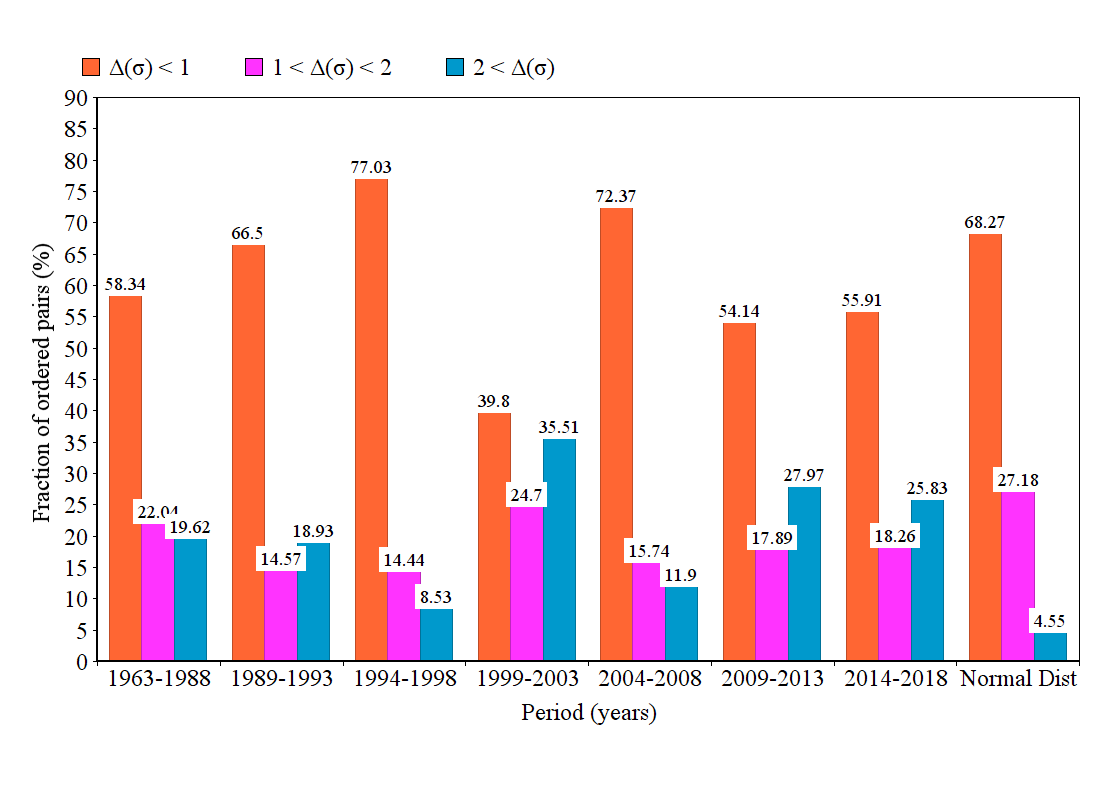}
 \caption{Percentage of ordered pairs of extragalactic distance moduli over ranges of $\Delta(\sigma)$ for different periods from 1963 to 2018, compared with a normal distribution. Too large fraction of the ordered pairs differ by a significantly large multiple of their reported uncertainties [$\Delta(\sigma)>2$], for distance moduli measurements published in the latest two periods 2009 to 2013 (27.97\%) and 2014 to 2018 (25.83\%), pointing to a possible increasing underestimation of uncertainties with time} %% no full stop at the end of caption
 \label{fig:6}
 \end{figure*}

Fig. \ref{fig:6} shows the percentage of ordered pairs of extragalactic distance moduli over ranges of $\Delta(\sigma)$ for each of the periods. A large fraction of the ordered pairs differ by a significantly large multiple of their reported uncertainties [$\Delta(\sigma)>2$] for distance moduli measurements published in both the periods 2009 to 2013 (27.97\%) and 2014 to 2018 (25.83\%), with the fraction increasing from 2004 to 2018. The fractions are much greater than the expected percentage for a normal distribution (4.55\%), and point to a possible increasing underestimation of uncertainties with time.

\begin{figure*}
 \centering\includegraphics[width=0.75\textwidth]{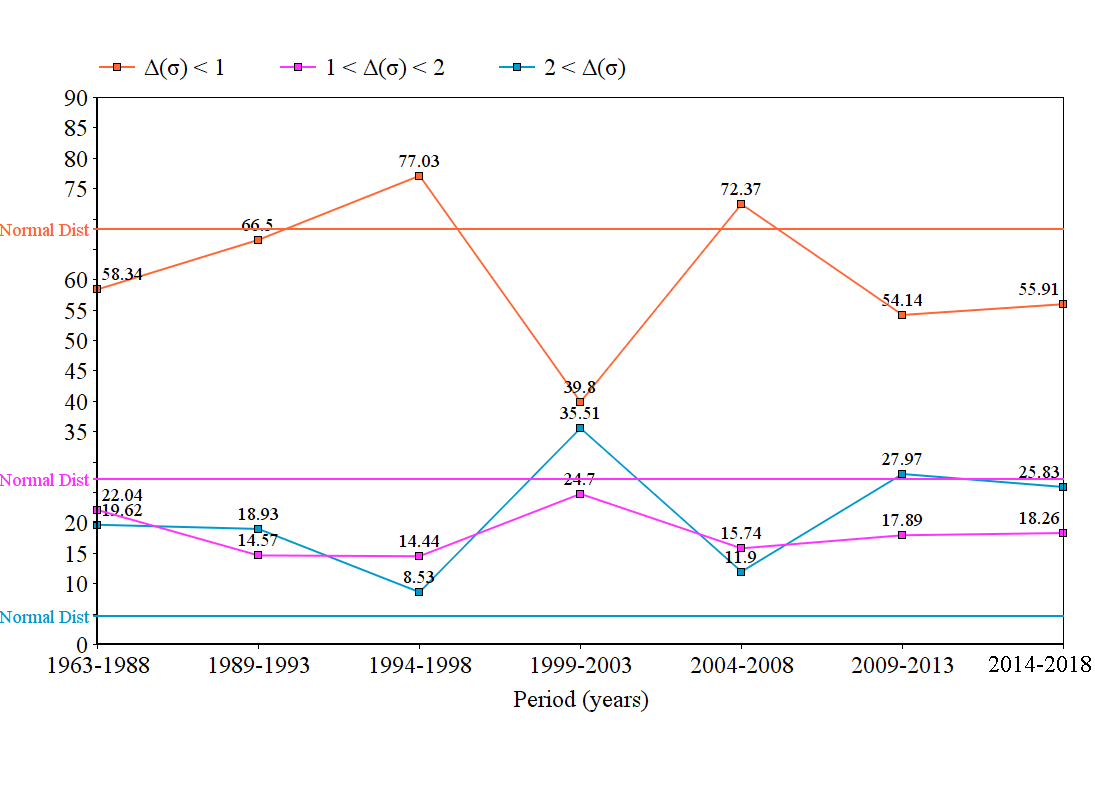}
 \caption{Variation of percentage of ordered pairs over ranges of $\Delta(\sigma)$ with period of publication, compared with a normal distribution. In the latest two periods 2009 to 2013 and 2014 to 2018, a large fraction of the ordered pairs of distance moduli differ by a significantly large multiple of their reported uncertainties [$\Delta(\sigma)>2$],  with the fraction increasing from 2004 to 2018, pointing to a possible increasing underestimation of uncertainties with time} %% no full stop at the end of caption
 \label{fig:7}
 \end{figure*}

 Fig. \ref{fig:7} shows the variation of percentage of ordered pairs over ranges of $\Delta(\sigma)$ with period of publication. For distance moduli reported in the latest two periods 2009 to 2013 and 2014 to 2018, a large fraction of the ordered pairs differ by a significantly large multiple of their reported uncertainties [$\Delta(\sigma)>2$], with the fraction increasing from 2004 to 2018. This points to a possible increasing underestimation of uncertainties with time.
 
 \section{Conclusion}
 
 This paper introduces a metric to assess the statistical self-consistency of reported uncertainties of extragalactic distance measurements. By comparing this measure of consistency of uncertainties over different distance measurement methods, and over time, we find some evidence for possible systematic underestimation of uncertainties throughout the distance ladder, and some evidence for increasing underestimation of uncertainties over the decades.

A large fraction of extragalactic distances report so low uncertainties that the mean difference between measured distances of the same galaxy across the NED database is more than twice the reported uncertainty. Tables \ref{tbl:3} and \ref{tbl:4} in the appendix, and Figures \ref{fig:3} and \ref{fig:4} in the article  show that this occurs across a large number of distance measurement methods. It is particularly pronounced in Standard Candles and Standard Rulers, as can be seen from Table 1 and Figure 2. This strongly points to a possible underestimation of uncertainties in extragalactic distances.

Furthermore, this trend has worsened with time. Table \ref{tbl:2} and Fig. \ref{fig:5} show that the mean reported one sigma uncertainty (mean $\sigma$) has markedly decreased from 2004 to 2018, while the average difference between two reported distances to the same galaxy as a multiple of the reported uncertainty [mean $\Delta(\sigma)$] has significantly increased from 2004 to 2018.

It is striking to note that this problem of underestimation of uncertainties in extragalactic distances has been known since 1985, and yet, it continues to worsen \citep{rowan-robinson_cosmological_1985, jacoby_critical_1992}.

Therefore, we conclude that the possibility of systematic underestimation of uncertainties in extragalactic distances should be explored. More liberal estimation of uncertainties in future reported extragalactic distances should be considered. Impact of higher uncertainties in extragalactic distances on our confidence in cosmological parameters and models should be evaluated.

For extragalactic distances measured using Type Ia Supernovae, we find that any two distance measurements of the same galaxy differ, on average, by 2.85 times the reported uncertainty in any of them. Since Type Ia Supernovae distances form a critical basis of cosmological models and expansion history, underestimation of uncertainties in them may have some implications for our confidence in cosmological parameters and models.

The results also give a possible way out of the Hubble-Lemaitre tension by advocating for increasing the error bars on Hubble-Lemaitre constant measured via distance ladder of standard candles and rulers. If this is done, the dichotomy between the values of the Hubble-Lemaitre constant, estimated from observations of the CMB, and those estimated from Type Ia Supernovae and TRGB observations, ceases to exist.

We humbly believe that the entire academic community—researchers, supervisors, reviewers, funders, and publishers—should work to reduce the pressure of reducing existing uncertainties, and even a reconfirmation of an existing measurement should be accorded greater significance. 

With deeply felt gratitude to the anonymous Reviewer 1, we reproduce their powerful one sentence summary of the entire article: “The pervasive attitude that smaller error bars imply a morally superior measurement is a sociological cause of the effects quantified in this paper.”

%\subsection{}%\label{ss:?}
%\subsubsection{}%\label{sss:?}

%% Math 
%
%\begin{eqnarray}%\label{eqn:?}
%\\ \nonumber
%\end{eqnarray}
%
%\begin{equation}%\label{eqn:?}
%\end{equation}

%% Table (two-column)
%
% \begin{table*}%%[tb]
% \small
% \caption{Caption} %% no full stop at the end of caption
%  \label{tbl:?}
% \begin{tabular}{}
% \tableline  %% rule at top
% \tablenotemark{a} 
% <entries>
% \tableline %% rule at bottom
% \end{tabular}
%
%% Any table notes must follow the \end{tabular} command. 
% \tablenotetext{a}{}
% \tablecomments{}
% \end{table*}

%% Table (one-colum)
%
% \begin{table}
% \caption{} %% no full stop at the end of caption
% \label{tbl:?}
% \begin{tabular}{}
% \tableline  %% rule at top
% \tablenotemark{a} 
% <entries>
% \tableline %% rule at bottom
% \end{tabular}
% \end{table}

%% Deluxe tabel (refer to AASTeX documentation)
%
% \begin{deluxetable}{ccrrrrrrrrcrl}
% \tabletypesize{\scriptsize}
% \rotate
% \tablecaption{}% %% no full stop at the end of caption
% \label{tbl:?} 
% \tablewidth{0pt}
% \tablehead{\colhead{}}
% \startdata
% <entries>
% \enddata
%
%% Text for table notes should follow after the \enddata but before
%% the \end{deluxetable}. Make sure there is at least one \tablenotemark
%% in the table for each \tablenotetext.
% \tablecomments{}
% \tablenotetext{a}{}
% \tablenotetext{b}{}
% \end{deluxetable}

%% Figure 
%
% \begin{figure}%[tb]
% \includegraphics{}
% \includegraphics[width=\columnwidth]{}
% \caption{} %% no full stop at the end of caption
% \label{fig:?}
% \end{figure}

%% Acknowledgements
%
 \acknowledgments
 We thank Stephan Kolassa, Semoi, and James K on stackexchange.com. We deeply thank Ian Steer, the anonymous Reviewer 1, and the Editor, who were more impactful to this work than the author. We also thank the Reviewers 2 and 3 for their most kind guidance in improving the article.
% <Acnowledgments text>

%% Available additional data environments: 
%% authorcontribution, fundinginformation, dataavailability, materialsavailability, codeavailability.
% \begin{authorcontribution}
% <some text> 
% \end{authorcontribution}

\begin{dataavailability}
Some of the data underlying this article are available in the NED Redshift-Independent Distances database at http://ned.ipac.caltech.edu/Library/Distances/. Rest of the data underlying this article are available in the article and the appendix.
\end{dataavailability}

\begin{codeavailability}
The code is released open source under the MIT license at https://github.com/RiteshSingh/galaxydistance.
\end{codeavailability}

 \begin{ethics}
 \begin{conflict}
 None.
% <conflict of interest declaration>
 \end{conflict}
 \begin{fundinginformation}
 None.
 \end{fundinginformation}
 \end{ethics}

%% References
%% Please cite all reference entries in the article text using \cite or
%% equivalent command. 

%%%  Using BibTeX  (Name-Year style)
%
\bibliographystyle{spr-mp-nameyear-cnd}  %% BibTeX style
\bibliography{citations}                %% BibTeX data

%% Non-BibTeX  (Name-Year style)
%
% \begin{thebibliography}{}
% \bibitem[\protect\citeauthoryear{<author>}{<year>]{ref:?}
%    <ref. entry>
% \bibitem[\protect\citeauthoryear{<author>}{<year>]{ref:?}
%    <ref. entry>
% \end{thebibliography}

\appendix
%\textbf{Appendix: Complete analysis results for each distance measurement method separately}
 
  %% Table (two-column)
%
 \begin{table*}%%[tb]
 \small
 \caption{Mean reported one sigma uncertainty and mean $\Delta(\sigma)$ for 44 distance measurement methods with multiple reported measurements for at least one galaxy in the NED database \citep{steer_redshift-independent_2017}} %% no full stop at the end of caption
 \label{tbl:3}
 \begin{tabular}{c|c|c|c|c|c}
 \tableline  %% rule at top
Distance method & Mean $\sigma$ (\%) & No. of observations & Mean $\Delta(\sigma)$ & No. of pairs & No. of galaxies \\ \hline
\emph{Standard Candles} \\ \hline
AGN Time Lag & 1.05 & 23 & \color{red}9.39 & 26 & 11 \\
BL Lac Luminosity & 1.16 & 5 & 0.97 & 8 & 2 \\
Brightest Stars & 1.85 & 62 & \color{red}2.45 & 238 & 19 \\
Carbon Stars & 2.73 & 33 & 0.42 & 338 & 6 \\
Cepheids & 0.40 & 1321 & \color{red}4.26 & 47684 & 61 \\
CM Diagrams & 0.63 & 325 & 1.47 & 15152 & 63 \\
Gamma-Ray Burst & 1.84 & 639 & 1.69 & 3264 & 136 \\
GCLF & 0.76 & 748 & \color{red}2.67 & 4464 & 176 \\
GCSBF & 0.49 & 2 & 0.33 & 2 & 1 \\
Horizontal Branch & 0.70 & 61 & 1.38 & 168 & 19 \\
Miras & 0.92 & 14 & 1.35 & 28 & 5 \\
Novae & 1.03 & 13 & 0.76 & 20 & 5 \\
PNLF & 0.60 & 147 & 1.19 & 412 & 43 \\
RR Lyrae Stars & 0.53 & 276 & \color{red}2.26 & 3318 & 41 \\
Red Clump & 0.50 & 32 & \color{red}2.19 & 142 & 7 \\
Red Supergiants & 0.54 & 5 & 0.66 & 6 & 2 \\
SNIa SDSS & 0.69 & 2481 & \color{red}2.41 & 2584 & 1225 \\
SX Phoenicis Stars & 1.75 & 2 & 0.22 & 2 & 1 \\
Statistical & 0.35 & 31 & \color{red}3.38 & 52 & 12 \\
Sunyaev-Zeldovich & 1.37 & 280 & \color{red}4.65 & 836 & 78 \\
SBF & 0.58 & 1648 & 1.76 & 8254 & 406 \\
TRGB & 0.44 & 1539 & \color{red}2.82 & 16154 & 305 \\
Type II Cepheids & 0.55 & 20 & 0.72 & 126 & 4 \\
Type Ia Supernovae & 0.48 & 14888 & \color{red}2.85 & 124016 & 2518 \\ \hline
\emph{Standard Rulers} \\ \hline
Dwarf Galaxy & 2.09 & 16 & 0.22 & 16 & 8 \\
Eclipsing Binary & 0.66 & 13 & 1.60 & 54 & 3 \\
GC Radii & 0.51 & 2 & 0.53 & 2 & 1 \\
Gravitational Lens & 2.03 & 48 & \color{red}2.31 & 140 & 15 \\
HII Region & 0.51 & 35 & \color{red}3.22 & 36 & 17 \\
Masers & 1.12 & 14 & \color{red}3.17 & 44 & 4 \\
Ring Diameter & 2.41 & 94 & 0.53 & 86 & 47 \\
Type II Supernovae & 0.74 & 747 & \color{red}2.24 & 4830 & 149 \\ \hline
\emph{Secondary Methods} \\ \hline
D-Sigma & 1.03 & 1974 & 1.09 & 5786 & 551 \\
Diameter & 1.45 & 6 & 0.48 & 6 & 3 \\
Faber-Jackson & 1.86 & 1415 & 1.07 & 3818 & 427 \\
Fundamental Plane & 0.88 & 2163 & 1.48 & 3274 & 966 \\
GeV TeV Ratio & 1.88 & 10 & 1.74 & 16 & 4 \\
GCFP & 0.97 & 3 & \color{red}2.71 & 6 & 1 \\
Gravitational Wave & 1.09 & 2 & 0.56 & 2 & 1 \\
IR Astro Satellite & 2.39 & 913 & 0.03 & 786 & 405 \\
Mass Model & 2.93 & 5 & \color{red}4.91 & 8 & 2 \\
Tertiary & 0.63 & 2 & \color{red}4.22 & 2 & 1 \\
Tully Estimate & 2.50 & 6 & 0.04 & 6 & 3 \\
Tully-Fisher & 1.22 & 54134 & 0.94 & 459656 & 9105 \\
% \tablenotemark{a} 
% <entries>
 \tableline %% rule at bottom
 \end{tabular}
%
%% Any table notes must follow the \end{tabular} command. 
% \tablenotetext{a}{}
 \tablecomments{The values with mean $\Delta(\sigma)>2$ corresponding to a p-value <4.55\% are colored red, pointing to a possible underestimation of uncertainties.}
 \end{table*}
 
  %% Table (two-column)
%
 \begin{table*}%%[tb]
 \small
 \caption{Percentage of ordered pairs of extragalactic distance moduli over ranges of $\Delta(\sigma)$, for each of the 44 distance measurement methods with multiple reported measurements for at least one galaxy in the NED database \citep{steer_redshift-independent_2017}} %% no full stop at the end of caption
 \label{tbl:4}
 \begin{tabular}{c|c|c|c|c}
 \tableline  %% rule at top
Distance method & $\Delta(\sigma)<1$ (\%) & $1<\Delta(\sigma)<2$ (\%) & $2<\Delta(\sigma)$ (\%) & No. of galaxies \\ \hline
\emph{Standard Candles} \\ \hline
AGN Time Lag & 19.23 & 15.38 & \color{red}65.39 & 11 \\
BL Lac Luminosity & 75.00 & 12.50 & 12.50 & 2 \\
Brightest Stars & 42.86 & 34.45 & 22.69 & 19 \\
Carbon Stars & 95.56 & 3.55 & 0.89 & 6 \\
Cepheids & 29.77 & 22.13 & \color{red}48.10 & 61 \\
CM Diagrams & 52.11 & 28.47 & 19.42 & 63 \\
Gamma-Ray Burst & 59.34 & 21.29 & 19.37 & 136 \\
GCLF & 35.03 & 24.22 & \color{red}40.75 & 176 \\
GCSBF & 100.00 & 0.00 & 0.00 & 1 \\
Horizontal Branch & 51.79 & 25.00 & 23.21 & 19 \\
Miras & 50.00 & 25.00 & \color{red}25.00 & 5 \\
Novae & 80.00 & 15.00 & 5.00 & 5 \\
PNLF & 58.98 & 26.70 & 14.32 & 43 \\
RR Lyrae Stars & 45.18 & 21.76 & \color{red}33.06 & 41 \\
Red Clump & 55.63 & 21.83 & 22.54 & 7 \\
Red Supergiants & 66.67 & 33.33 & 0.00 & 2 \\
SNIa SDSS & 32.00 & 23.88 & \color{red}44.12 & 1225 \\
SX Phoenicis Stars & 100.00 & 0.00 & 0.00 & 1 \\
Statistical & 30.77 & 28.85 & \color{red}40.38 & 12 \\
Sunyaev-Zeldovich & 46.89 & 18.66 & \color{red}34.45 & 78 \\
SBF & 47.60 & 24.23 & \color{red}28.17 & 406 \\
TRGB & 40.72 & 22.17 & \color{red}37.11 & 305 \\
Type II Cepheids & 78.57 & 17.46 & 3.97 & 4 \\
Type Ia Supernovae & 41.82 & 20.66 & \color{red}37.52 & 2518 \\ \hline
\emph{Standard Rulers} \\ \hline
Dwarf Galaxy & 100.00 & 0.00 & 0.00 & 8 \\
Eclipsing Binary & 55.55 & 18.52 & \color{red}25.93 & 3 \\
GC Radii & 50.00 & 50.00 & 0.00 & 1 \\
Gravitational Lens & 47.86 & 15.71 & \color{red}36.43 & 15 \\
HII Region & 44.44 & 16.67 & \color{red}38.89 & 17 \\
Masers & 38.64 & 34.09 & \color{red}27.27 & 4 \\
Ring Diameter & 88.37 & 9.30 & 2.33 & 47 \\
Type II Supernovae & 44.05 & 21.00 & \color{red}34.95 & 149 \\ \hline
\emph{Secondary Methods} \\ \hline
D-Sigma & 69.29 & 19.18 & 11.53 & 551 \\
Diameter & 100.00 & 0.00 & 0.00 & 3 \\
Faber-Jackson & 58.30 & 28.65 & 13.05 & 427 \\
Fundamental Plane & 57.21 & 20.46 & 22.33 & 966 \\
GeV TeV Ratio & 43.75 & 25.00 & \color{red}31.25 & 4 \\
GC Fundamental Plane & 50.00 & 16.67 & \color{red}33.33 & 1 \\
Gravitational Wave & 100.00 & 0.00 & 0.00 & 1 \\
IR Astro Satellite & 99.87 & 0.13 & 0.00 & 405 \\
Mass Model & 25.00 & 0.00 & \color{red}75.00 & 2 \\
Tertiary & 0.00 & 0.00 & \color{red}100.00 & 1 \\
Tully Estimate & 100.00 & 0.00 & 0.00 & 3 \\
Tully-Fisher & 68.77 & 21.01 & 10.22 & 9105 \\ \hline
Normal Distribution & 68.27 & 27.18 & 4.55 \\
% \tablenotemark{a} 
% <entries>
 \tableline %% rule at bottom
 \end{tabular}
%
%% Any table notes must follow the \end{tabular} command. 
% \tablenotetext{a}{}
 \tablecomments{The values with $\Delta(\sigma)>2$ for more than 25\% of the ordered distance measurement pairs are colored red, pointing to a possible underestimation of uncertainties.}
 \end{table*}

\end{document}